\documentclass[12pt]{article}
\usepackage{blindtext}
\usepackage{scrextend}
\usepackage{hyperref}
\usepackage{amsmath}
\usepackage{amsfonts}
\usepackage{amssymb}
\usepackage{amsthm}                                                                                                                                                                                                                                        
\usepackage{times}
\usepackage{MnSymbol}

\newcommand{\ket}[1]{| #1 \rangle}

\title{Understanding the\\ Frauchiger-Renner Argument}
\author{Jeffrey Bub\\ \small Philosophy Department\\\small Institute for Physical Science and Technology\\\small Joint Center for Quantum Information and Computer Science\\  \small University of Maryland, College Park, MD 20742, USA}

\date{}
\normalsize
\begin{document}
\maketitle

\begin{abstract}
In 2018, Daniela Frauchiger and Renato Renner published an article  in \emph{Nature Communications}  entitled `Quantum theory cannot consistently describe the use of itself.' I clarify the significance of the result and point out a common and persistent misunderstanding of the argument, which has been attacked as flawed from a variety of interpretational perspectives.
\end{abstract}

In 2018, Daniela Frauchiger and Renato Renner published an article  in \emph{Nature Communications}  entitled `Quantum theory cannot consistently describe the use of itself' \cite{Frauchiger+2018}. Here I want to point out a common misunderstanding of the Frauchiger-Renner argument, which has been attacked as flawed from a variety of interpretational perspectives.

As their Abstract states, Frauchiger and Renner propose `a Gedankenexperiment to investigate the question whether quantum theory can, in principle, have universal validity.' Specifically, the issue is whether it is possible `to employ quantum theory to model complex systems that include agents who are themselves using quantum theory.' The claim is that, given certain assumptions, the agents' conclusions derived from quantum theory are inconsistent: `one agent, upon observing a particular measurement outcome, must conclude that another agent has predicted the opposite outcome with certainty.'

The Gedankenexperiment is a variant of Wigner's experiment in the well-known Wigner's Friend argument \cite{Wigner1961}, modified by a construction by Hardy \cite{Hardy2000,Hardy1993}. In Wigner's experiment, the Friend, $F$, measures the $z$-spin of a qubit in an isolated laboratory $L$ containing the qubit  $Q$, the measuring apparatus $A$, and $F$. After the measurement, the laboratory is in an entangled state $\ket{\Psi}_{L}$, a linear superposition of product states of the form $\ket{+}_{L} = \ket{+\frac{1}{2}}_{Q}\ket{\mbox{`+'}}_{A}\ket{\mbox{``+''}}_{F}$ and $\ket{-}_{L} = \ket{-\frac{1}{2}}_{Q}\ket{\mbox{`-'}}_{A}\ket{\mbox{``-''}}_{F}$, where $\ket{\mbox{`+'}}_{A}\ket{\mbox{``+''}}_{F}$ and $\ket{\mbox{`-'}}_{A}\ket{\mbox{``-''}}_{F}$ represents states of the apparatus recording the $+\frac{1}{2}$ or $-\frac{1}{2}$ outcome and $F$ registering this outcome. Wigner, a `super-observer' $W$ outside the laboratory, is assumed to have the technological ability to measure arbitrary observables of $L$, e.g., an observable with eigenstates $\ket{+}_{L}, \ket{-}_{L}$, or the observable represented by the projection operator onto the entangled state $\ket{\Psi}_{L}$. 

The Frauchiger-Renner Gedankenexperiment involves a timed sequence of measurements by $F$ and an additional Friend, $\overline{F}$, and $W$ and an additional Wigner, $\overline{W}$. In addition to measurements, the agents make inferences about `certainty' on the basis of the measurement outcomes according to two inference rules, $Q$ and $C$, and a third rule, $S$, which prohibits inconsistent inferences and is not invoked until the last step of the argument. 

Rule $Q$: If (i) an agent $A$ has established that a quantum system $Q$ is in a state $\ket{\psi}_{Q}$ at time $t_{0}$, and the Born probability of the outcome $\xi$ of a measurement of an observable $X$ on $Q$ in the state $\ket{\psi}$ completed at time $t$ is 1, or (ii) if an agent has observed the outcome $\xi$ of a measurement of $X$ on $Q$ completed at time $t$, then  agent $A$ can conclude: `I am certain that $x = \xi$ at time $t$.'

Rule $C$: If an agent $A$ has established: `I am certain that another agent $A'$, whose inferences about certainty are in accordance with $Q, C$, and $S$, is certain that  $x = \xi$ at time $t$,' then agent $A$ can conclude: `I am certain that $x = \xi$ at time $t$.'

Rule $S$: If an agent $A$ has established `I am certain that $x = \xi$ at time $t$,' then agent $A$ cannot also establish  `I am certain that $x \neq \xi$ at time $t$.'

I have amended the original version of rule $Q$, as stated in the Frauchiger-Renner article, by explicitly adding (ii), the inference to certainty after an agent has measured an observable to have a certain value. Renner agrees (private communication) that rule $Q$ should be understood as including (ii).

As Frauchiger and Renner point out \cite[p. 3]{Frauchiger+2018}, we should think of the agents as quantum computers programmed to carry out the sequence of measurements in the Gedankenexperiment, and to draw inferences about `certainty' from the measurement outcomes according to the inference rules $Q$ and $C$, constrained by the consistency requirement, rule $S$. Each agent stores the statements established as `certain' according to these rules in an internal memory register. Inferences to `certainty' are licensed by the three rules and nothing more, and the significance of `being certain' is implicit in the constraints defined by these rules and nothing more. As Renner put it (private communication), `being certain' can mean whatever you like, provided the use of the term is in accordance with the three inference rules. 

To understand the argument, it is crucial to see that the agents are modeled as \emph{physical} systems (quantum systems), and the reasoning processes of the agents are \emph{physical processes} that can occur in superposition. An agent isn't required to always draw a definite conclusion  but can draw conclusions in superposition.  In particular, an agent can be in a superposition of `being certain' and not `being certain,' or of `being certain' and drawing no conclusion. The goal here is to investigate the question whether it is possible `to employ quantum theory to model complex systems that include agents who are themselves using quantum theory.'  The rules are minimal assumptions about agents, modeled as quantum systems, who use quantum theory to make inferential moves that are represented as physical processes, i.e., unitary evolutions.

The experiment is repeated over many rounds, $n = 0, 1, 2, \ldots$ until a halting condition is satisfied. A round lasts from  $n:00$ to $n:40$, where the notation $n:xx$ represents a `time' in minutes ($n$) and seconds ($xx$). Here is the experimental protocol:

At  $n:00$, $\overline{F}$ invokes a randomness generator based on the measurement of a quantum system, $R$, initially in a state $\ket{\mbox{init}}_{R} = \sqrt{1/3}\ket{\mbox{heads}}_{R} + \sqrt{2/3}\ket{\mbox{tails}}_{R}$. She sets the spin of an electron, $S$, to $\ket{\downarrow}_{S}$ if $r = \mbox{heads}$ or to $\ket{\rightarrow}_{S} = \sqrt{1/2}(\ket{\downarrow}_{S} + \ket{\uparrow}_{S})$  if $r = \mbox{tails}$, and sends $S$ to $F$.

From $n:01$ to $n:04$, $\overline{F}$ uses  rule $Q$ to either establish as certain that $w = \mbox{fail}$ if the outcome of her  measurement is $r = \mbox{tails}$, which she stores in her memory register, or she draws no conclusion about the value $w$. 

 At $n:10$, $F$ measures $S$ with respect to the basis \{$\ket{\downarrow}_{S}, \ket{\uparrow}_{S}$\} and records the outcome $z \in \{-\frac{1}{2}, +\frac{1}{2}\}$.
 
 From $n:11$  to $n:14$, $F$ applies rule  $Q$ to establish as certain that $r = \mbox{tails}$ if the outcome of her spin measurement is $z = +\frac{1}{2}$, and then applies rule $C$ to establish as certain that $w = \mbox{fail}$, which she stores in her memory register, or she draws no conclusion about the value of $w$. 

At $n:20$,  $\overline{W}$ measures $\overline{L}$ with respect to a basis \{$\ket{\overline{\mbox{ok}}}_{\overline{L}} = \sqrt{1/2}(\ket{\overline{h}}_{\overline{L}} - \ket{\overline{t}}_{\overline{L}}), \ket{\overline{\mbox{fail}}}_{\overline{L}} = \sqrt{1/2}(\ket{\overline{h}}_{\overline{L}} + \ket{\overline{t}}_{\overline{L}})$\}. Here $\ket{\overline{h}}_{\overline{L}}$ is the state that $\overline{F}$'s laboratory $\overline{L}$ would evolve to at time $n:20$ if the initial state of $R$ is  $\ket{\mbox{heads}}_{R}$. This is the same as the state of $\overline{L}$ after $\overline{F}$ completes her measurement of $R$ and stores her inference in her internal memory at $n:04$, since nothing happens to $\overline{L}$ between $n:04$ and $n:20$. The state $\ket{\overline{t}}_{\overline{L}}$ is defined similarly. He records $\overline{w} = \overline{\mbox{ok}}$ or $\overline{w} = \overline{\mbox{fail}}$ depending on the outcome. The record in $\overline{W}$'s memory register is available  to outside observers as an announcement or display, and also accessible by $W$.

From $n:21$  to $n:24$, $\overline{W}$ applies rule $Q$ to establish as certain that $z = +\frac{1}{2}$ if the outcome of his measurement is  $\overline{w} = \overline{\mbox{ok}}$, and then applies rule $C$ to establish as certain that $w = \mbox{fail}$, which he stores in his memory register; he draws no conclusion about the value of $w$ if the outcome of his measurement is  $\overline{w} = \overline{\mbox{fail}}$. 

From $n:26$ to $n:29$, $W$ accesses $\overline{W}$'s memory and either draws no conclusion about the value of $w$ if $\overline{W}$ drew no conclusion or, if $\overline{W}$ established as certain that $w = \mbox{fail}$, $W$ records as certain that $w = \mbox{fail}$ using rule $C$ and stores the information in his memory register. 

At $n:30$, $W$ measures $L$ with respect to a basis \{$\ket{\mbox{ok}}_{L} = \sqrt{1/2}(\ket{-\frac{1}{2}}_{L} - \ket{+\frac{1}{2}}_{L}), \ket{\mbox{fail}}_{L} = \sqrt{1/2}(\ket{-\frac{1}{2}}_{L} + \ket{+\frac{1}{2}}_{L})$\}. The states $\ket{+\frac{1}{2}}_{L}$  and $\ket{-\frac{1}{2}}_{L}$  of $L$ are defined similarly to $\ket{\overline{h}}_{\overline{L}}$  and $\ket{\overline{t}}_{\overline{L}}$   and correspond to the states $\ket{+}_{L}$ and $\ket{-}_{L}$ in the above description of Wigner's original thought experiment.  The state $\ket{+\frac{1}{2}}_{L}$ is the state that $F$'s laboratory $L$ would evolve to at time $n:30$ if the initial state of $S$ is $\ket{\uparrow}_{S}$. This is the same as the state of $L$ after $F$ completes her spin measurement and stores her inference in her internal memory at $n:24$, since nothing happens to $L$ between $n:24$ and $n:30$. The state $-\frac{1}{2}$ is defined similarly. He records $w = \mbox{ok}$ or $w = \mbox{fail}$ depending on the outcome. The record in $W$'s memory register is available to outside observers.

From $n:31$ to $n:34$, $W$ checks for consistency using rule $S$.

At $n:40$, the experiment is halted if $\overline{w} = \overline{\mbox{ok}}$ and $w = \mbox{ok}$.

The whole point of the argument is to show that if agents, capable of performing measurements on quantum systems and making inferences on the basis of these measurements, are themselves modeled as quantum systems evolving unitarily in accordance with the rules $Q, C, S$, then, for the particular Gedankenexperiment considered, we end up with an inconsistency. So, it  is crucial to the argument that these computers and their memory registers are quantum systems that evolve unitarily at all times. 

Frauchiger and Renner  \cite[p. 2]{Frauchiger+2018} point to the circuit diagram in the Methods section as showing that each step of the experiment can be described by a unitary evolution on a particular subsystem. A widely circulated but unpublished Note by Renner \cite{Renner}  shows the unitary evolution in detail. With Renner's permission, I will follow the analysis in the Note.

Losada, Laura, and Lombardi \cite{Losada+2019} argue, using the `consistent histories' approach to quantum mechanics, that `the supposedly contradictory conclusion of the [Frauchiger-Renner] argument requires computing probabilities in a family of histories that does not satisfy the consistency condition, i.e., an invalid family of histories for the theory.' As the authors note on p. 3, `a history is defined as a sequence of events at different times, where an event is the occurrence of a property.' So what they show, in effect, is that an assignment of properties, or values to the observables of the experiment, corresponding to the steps in the argument, is not a consistent history. But this is irrelevant to the Frauchiger-Renner argument.

\emph{The Frauchiger-Renner argument does not assume, nor needs to assume, anything about the assignment of values to observables---in particular, the argument does not assume that observable values are relative to an agent. The entire argument is framed in terms of measurement outcomes, and what the agents can be `certain' about at various times according to the rules $Q, C$ and $S$, which is what the agents store in their memory registers.}

It is worth pointing out here that if an agent `is certain' that the value of an observable $X$ at time $t$ is $\xi$, it does not follow that the value of $X$ is $\xi$ at time $t$. What follows is just what can be inferred from the three inference rules, and nothing more. If an agent measures $X$ and finds the outcome $\xi$ at time $t$, the argument does not assume, and isn't required to assume, that the value of $X$ at time $t$ is $\xi$---rather, it follows only that the agent `is certain' that the value of $X$ is $\xi$ at time $t$. 

What comes to  much the same thing is the claim that the argument assumes that when an agent in an isolated laboratory performs a measurement, the state collapses inside the laboratory, but not outside. Muci\~{n}o and Okon \cite{Mucino+2020} argue that, in addition to the three rules, Frauchiger and Renner implicitly invoke the assumption:
\begin{quote}
(H) When a measurement is carried out inside of a closed lab, such a measurement leads to a collapse for inside observers, but it does not lead to a collapse for outside observers.
\end{quote}

\emph{The Frauchiger-Renner argument does not assume, nor needs to assume, anything other than a successive unitary evolution of the global quantum state through each stage of the experiment---in particular, the argument does not assume that the quantum state undergoes a `collapse' for observers inside a laboratory but not for outside observers.}

What might indeed require a collapse assumption is if agents were understood to always draw definite conclusions. But, as emphasized above, an agent is treated as a physical system and an agent's inferences are physical processes. So an agent can draw conclusions in superposition, and an agent can be in a superposition of `being certain' and not `being certain,' or of `being certain' and drawing no conclusion.

The first step in the experiment is $\overline{F}$'s  generation of the random number `heads' or `tails' at $n:00$ by measuring the system $R$ initially in the state $\ket{\mbox{init}}_{R} = \sqrt{2/3}\ket{\mbox{tails}}_{R} + \sqrt{1/3}\ket{\mbox{heads}}_{R}$, and $\overline{F}$'s subsequent preparation of the spin state of the electron $S$ in the state $\ket{\downarrow}_{S}$ or $\ket{\rightarrow}_{S}$, which she sends to $F$'s laboratory $L$. 

From $n:01$  to $n:04$,  the composite system consisting of $\overline{F}$'s laboratory and $S$ evolves to an entangled state, with one component corresponding to $\overline{F}$ finding `heads,' sending $S$ to $L$ in the state $\ket{\downarrow}_{S}$, and drawing no conclusion about $w$, and the other component corresponding to $\overline{F}$ finding `tails,' sending $S$ to $L$ in the state the state $\ket{\rightarrow}_{S}$ and inferring (establishing as certain) that $w = \mbox{fail}$ at $n:31$ (because after $F$'s spin measurement on $S$, this component will have evolved unitarily to the state $\ket{\mbox{fail}}$). Any action performed by $\overline{F}$ that has the effect of mapping orthogonal states to orthogonal states can be modeled as a unitary evolution, which in general produces an entangled state. In this case, $\overline{F}$'s measurement, followed by her  inference to  `$w$ = fail at $n:31$,' or a null inference to `no conclusion drawn,' can be modeled as a unitary evolution entangling the content of $\overline{F}$'s memory register with the output $r$ of the randomness generator and the spin state of $S$. The entangled state expresses a correlation, between the value of $r$, the prepared state of $S$, and $\overline{F}$'s prediction (either ``$w$ = fail at $n:31$,' or `no conclusion drawn').

Without loss of generality, we can assume that  the initial state of $\overline{F}$'s memory register is some ready state $\ket{\bot}_{\overline{F}}$ and the initial state of $S$ is $\ket{\uparrow}_{S}$. The unitary describing $\overline{F}$'s measurement completed at $n:01$, followed by her inference completed at $n:02$, is defined by its effect on orthogonal states:
\begin{eqnarray}
\ket{\mbox{tails}}_{R}\ket{\bot}_{\overline{F}}\ket{\uparrow}_{S} & \rightmapsto & \ket{\mbox{tails}}_{R}\ket{\mbox{\tiny ``$r =$ tails, so I am certain that $w$ = fail at  $n:31$''}}_{\overline{F}}\ket{\rightarrow}_{S} \nonumber\\
\ket{\mbox{heads}}_{R}\ket{\bot}_{\overline{F}}\ket{\uparrow}_{S} & \rightmapsto & \ket{\mbox{heads}}_{R}\ket{\mbox{\tiny ``$r =$ tails, so no conclusion drawn''}}_{\overline{F}}\ket{\downarrow}_{S}
\end{eqnarray}
Applying this unitary results in the transition
\begin{eqnarray}
\ket{\mbox{init}}_{R}\ket{\bot}_{\overline{F}}\ket{\uparrow}_{S} &\rightmapsto& 
\ket{\Psi}^{n:02} = \sqrt{2/3}\ket{\mbox{tails}}_{R}\ket{\mbox{\tiny ``$r = $ tails, so I am certain that $w = $ fail at $n:31$''}}_{\overline{F}}\ket{\rightarrow}_{S} \nonumber \\
&&+ \sqrt{1/3}\ket{\mbox{heads}}_{R}\ket{\mbox{\tiny ``$r = $ heads, so no conclusion drawn''}}_{\overline{F}}\ket{\downarrow}_{S} \label{unitary1}
\end{eqnarray}

To see that $\overline{F}$'s inference to `$w = $ fail' is correct, consider the isometry $U$ from $S$ to $L$ that describes the unitary evolution between the preparation of the spin state of $S$ and $W$'s measurement of $L$. This isometry maps $\ket{\rightarrow}_{S}$ to $\ket{\mbox{fail}}_{L}$. The inverse, $U^{-1}$, maps $\ket{\mbox{fail}}_{L}$ back to $\ket{\rightarrow}_{S}$. The state $\ket{\Psi}^{n:02}$ can then be expressed as
\[
\ket{\Psi}^{n:02} = \sqrt{2/3}\ket{\mbox{tails}}_{R}\ket{\mbox{\tiny ``$r = $ tails, so I am certain that $w = $ fail at $n:31$''}}_{\overline{F}}U^{-1}\ket{\mbox{fail}}_{L} + \sqrt{1/3}\,\cdots
\]
One can now read off directly from this state that, in the component where $\overline{F}$ does draw a conclusion, this conclusion agrees with what is the case in $L$.

The global state, initially a product state $\ket{\Psi}^{n:00} = \ket{\mbox{init}}_{R}\ket{\bot}_{\overline{F}}\ket{\uparrow}_{S}\ket{\bot}_{F}\ket{\bot}_{\overline{W}}\ket{\bot}_{W}$ at the beginning of each round $n$ of the experiment, evolves unitarily to the entangled  state $\ket{\Psi}^{n:02}$ on the basis of transformation rules intended to model measurements and inferential moves following the registration of a measurement outcome by a human agent using quantum theory. The sequence of subsequent measurements and inferences can all be modeled as unitary transformations in a similar way, as Renner shows in \cite{Renner}. The following is a close summary of Renner's analysis.  Only the relevant entangled component is explicitly indicated at each stage.

Consider $F$'s reasoning on the basis of observing $z = -\frac{1}{2}$ or $z = +\frac{1}{2}$. If $r = \mbox{heads}$, $\overline{F}$ prepares $S$ in the state $\ket{\downarrow}_{S}$. In that case, $F$'s measurement will yield $z = -\frac{1}{2}$. So if $F$ observes $z = +\frac{1}{2}$, she concludes via rule $Q$ that $r \neq \mbox{heads}$, i.e., $r = \mbox{tails}$. If she observes $z = -\frac{1}{2}$, she can't draw any conclusion about $r$. This inference can be characterized by the following unitary, which acts on $F$'s internal memory register, $F_{M}$:
\begin{eqnarray}
\ket{\mbox{\tiny ``I observed $z = +\frac{1}{2}$''}}_{F}\ket{\bot}_{F_{M}} & \rightmapsto & \ket{\mbox{\tiny ``I observed $z = +\frac{1}{2}$''}}_{F}\ket{\mbox{\tiny ``I am certain that $r =$ tails''}}_{F_{M}} \nonumber \\
\ket{\mbox{\tiny ``I observed $z = +-\frac{1}{2}$''}}_{F}\ket{\bot}_{F_{M}} & \rightmapsto & \ket{\mbox{\tiny ``I observed $z = -\frac{1}{2}$''}}_{F}\ket{\mbox{\tiny ``no conclusion drawn''}}_{F_{M}}
\end{eqnarray}
which can be expressed more simply as
\begin{eqnarray}
\ket{\mbox{\tiny ``I observed $z = +\frac{1}{2}$''}}_{F} & \rightmapsto & \ket{\mbox{\tiny ``I observed $z = +\frac{1}{2}$, so I am certain that $r =$ tails''}}_{F} \nonumber \\
\ket{\mbox{\tiny ``I observed $z = -\frac{1}{2}$''}}_{F} & \rightmapsto & \ket{\mbox{\tiny ``I observed $z = -\frac{1}{2}$;  no conclusion drawn''}}_{F}
\end{eqnarray}
or simply for brevity as
\begin{eqnarray}
\ket{\mbox{\tiny ``$z = +\frac{1}{2}$''}}_{F} & \rightmapsto & \ket{\mbox{\tiny ` $z = +\frac{1}{2}$, so I am certain that $r =$ tails''}}_{F} \nonumber \\
\ket{\mbox{\tiny ``$z = -\frac{1}{2}$''}}_{F} & \rightmapsto & \ket{\mbox{\tiny ``$z = -\frac{1}{2}$;  no conclusion drawn''}}_{F}
\end{eqnarray}

$F$'s inference  between $n:11$ and $n:12$ is described by this unitary, resulting in the transition to the state
\begin{eqnarray}
\ket{\Psi}^{n:12} & = & \sqrt{1/3}\ket{\mbox{tails}}_{R}\ket{\mbox{\tiny ``$r =$ tails, so I am certain that $w = $ fail at $n:31$''}}_{\overline{F}}\ket{\uparrow}_{S}\ket{\mbox{\tiny ``$z = +\frac{1}{2}$, so I am certain that $r =$ tails''}}_{F} \nonumber\\
&& + \sqrt{1/3}\ket{\mbox{tails}}_{R}\ket{\mbox{\tiny ``$r =$ tails, so I am certain that $w = $ fail at $n:31$''}}_{\overline{F}}\ket{\downarrow}_{S}\ket{\mbox{\tiny ``$z = -\frac{1}{2}$; no conclusion drawn''}}_{F} \nonumber \\
&& + \sqrt{1/3}\ket{\mbox{heads}}_{R}\ket{\mbox{\tiny ``$r =$ heads, so no conclusion drawn''}}_{\overline{F}}\ket{\downarrow}_{S}\ket{\mbox{\tiny ``$z = -\frac{1}{2}$; no conclusion drawn''}}_{F} 
\end{eqnarray}
A further unitary, implementing an inference via rule $C$ from $\overline{F}$'s inference about $w$ at $n:31$ to $F$'s certainty that $w = $ fail at $n:31$ or a null inference to `no conclusion drawn,' results in the state 
\begin{eqnarray}
\ket{\psi}^{n:14} & = & \sqrt{1/3}\ket{\mbox{tails}}_{R}\ket{\mbox{\tiny ``$r =$ tails, so I am certain that $w = $ fail at $n:31$''}}_{\overline{F}}\ket{\uparrow}_{S}\ket{\mbox{\tiny ``$z = +\frac{1}{2}$, so I am certain that $w = $ fail at $n:31$''}}_{F} \nonumber\\
&&+ \sqrt{1/3}\ket{\mbox{tails}}_{R}\ket{\mbox{\tiny ``$r =$ tails, so I am certain that $w = $ fail at $n:31$''}}_{\overline{F}}\ket{\downarrow}_{S}\ket{\mbox{\tiny ``I observed $z = -\frac{1}{2}$; no conclusion drawn''}}_{F} \nonumber \\
&&+ \sqrt{1/3}\ket{\mbox{heads}}_{R}\ket{\mbox{\tiny ``$r =$ heads, so no conclusion drawn''}}_{\overline{F}}\ket{\downarrow}_{S}\ket{\mbox{\tiny ``I observed $z = -\frac{1}{2}$; no conclusion drawn''}}_{F} 
\end{eqnarray}

Continuing in this way, and transforming to the basis \{$\ket{\overline{\mbox{ok}}}, \ket{\overline{\mbox{fail}}}$\} of $\overline{L}$ defined in the experimental protocol above, we find that after $\overline{W}$'s measurement and inferences from the outcome of his measurement (to $F$'s certainty about  $z$ via rule $Q$, and hence from $F$'s certainty about $w$ at $n:31$ to $\overline{W}$'s certainty about $w$ at $n:31$ via rule $C$, or to `no conclusion drawn'), the global state has the form
\begin{eqnarray}
\ket{\Psi}^{n:24} & = & \sqrt{1/6}\ket{\overline{\mbox{fail}}}_{\overline{L}}\ket{\mbox{\tiny ``$\overline{w} = \overline{\mbox{fail}}$; no conclusion drawn''}}_{\overline{W}}\ket{\uparrow}_{S}\ket{\mbox{\tiny ``$z = +\frac{1}{2}$, so I am certain that $w = $ fail at $n:31$''}}_{F} \nonumber \\
&& -\sqrt{1/6}\ket{\overline{\mbox{ok}}}_{\overline{L}}\ket{\mbox{\tiny ``$\overline{w} = \overline{\mbox{ok}}$, so I am certain that $w =$ fail at $n:31$''}}_{\overline{W}}\ket{\uparrow}_{S}\ket{\mbox{\tiny ``$z = +\frac{1}{2}$, so I am certain that $w = $ fail at $n:31$''}}_{F} \nonumber \\
&& +\sqrt{2/3}\ket{\overline{\mbox{fail}}}_{\overline{L}}\ket{\mbox{\tiny ``$\overline{w} = \overline{\mbox{ok}}$; no conclusion drawn''}}_{\overline{W}}\ket{\downarrow}_{S}\ket{\mbox{\tiny ``$z = -\frac{1}{2}$; no conclusion drawn''}}_{F} 
\end{eqnarray}

Note that while $F$'s certainty (at $n:14$) that $w =$ fail at $n:31$ relies on $\overline{F}$'s certainty (at $n:02$), and $\overline{F}$'s memory is erased by $\overline{W}$'s measurement of $\overline{L}$ at $n:20$, $\overline{W}$ relies on $F$'s memory to infer certainty that $w =$ fail at $n:31$, not $\overline{F}$'s memory, and $F$'s memory is not altered by $\overline{W}$'s measurement. 

Just before $n:30$, the global state in the basis \{$\ket{\mbox{ok}}, \ket{\mbox{fail}}$\} of $L$, defined in the experimental protocl,  can be expressed as
\begin{eqnarray}
\ket{\Psi}^{n:25} & = & \sqrt{1/12}\ket{\overline{\mbox{fail}}}_{\overline{L}}\ket{\mbox{\tiny ``$\overline{w} = \overline{\mbox{fail}}$; no conclusion drawn''}}_{\overline{W}}\ket{\mbox{fail}}_{L} \nonumber \\
&& -\sqrt{1/12}\ket{\overline{\mbox{fail}}}_{\overline{L}}\ket{\mbox{\tiny ``$\overline{w} = \overline{\mbox{fail}}$; no conclusion drawn''}}_{\overline{W}}\ket{\mbox{ok}}_{L} \nonumber \\ 
&& -\sqrt{1/12}\ket{\overline{\mbox{ok}}}_{\overline{L}}\ket{\mbox{\tiny ``$\overline{w} = \overline{\mbox{ok}}$, so I am certain that $w = $ fail at $n:31$''}}_{\overline{W}}\ket{\mbox{fail}}_{L} \nonumber \\
&& +\sqrt{1/12}\ket{\overline{\mbox{ok}}}_{\overline{L}}\ket{\mbox{\tiny ``$\overline{w} = \overline{\mbox{ok}}$, so I am certain that $w = $ fail at $n:31$''}}_{\overline{W}}\ket{\mbox{ok}}_{L} \nonumber \\
&& + \sqrt{1/3}\ket{\overline{\mbox{fail}}}_{\overline{L}}\ket{\mbox{\tiny ``$\overline{w} = \overline{\mbox{fail}}$; no conclusion drawn''}}_{\overline{W}}\ket{\mbox{fail}}_{L} \nonumber \\
&& +\sqrt{1/3}\ket{\overline{\mbox{fail}}}_{\overline{L}}\ket{\mbox{\tiny ``$\overline{w} = \overline{\mbox{fail}}$; no conclusion drawn''}}_{\overline{W}}\ket{\mbox{ok}}_{L}  
\end{eqnarray}

At $n:27$, $W$ accesses $\overline{W}$'s memory to learn the contents, and at $n:28$ he uses rule $C$ to infer `I am certain that $w$ = fail at $n:31$,' or the null inference `no conclusion drawn,' which he writes in his internal memory, resulting in a unitary evolution to the state
\begin{eqnarray}
\ket{\Psi}^{n:28} & = & \sqrt{1/12}\ket{\overline{\mbox{fail}}}_{\overline{L}}\ket{\mbox{\tiny ``$\overline{w} = \overline{\mbox{fail}}$; no conclusion drawn''}}_{\overline{W}}\ket{\mbox{fail}}_{L}\ket{\tiny\mbox{``no conclusion drawn''}}_{W} \nonumber \\
&& -\sqrt{1/12}\ket{\overline{\mbox{fail}}}_{\overline{L}}\ket{\mbox{\tiny ``$\overline{w} = \overline{\mbox{fail}}$; no conclusion drawn''}}_{\overline{W}}\ket{\mbox{ok}}_{L}\ket{\tiny\mbox{``no conclusion drawn''}}_{W} \nonumber \\ 
&& -\sqrt{1/12}\ket{\overline{\mbox{ok}}}_{\overline{L}}\ket{\mbox{\tiny ``$\overline{w} = \overline{\mbox{ok}}$, so I am certain that $w = $ fail at $n:31$''}}_{\overline{W}}\ket{\mbox{fail}}_{L}\ket{\tiny\mbox{``I am certain that $w =$ fail at $n:31$''}}_{W} \nonumber \\
&& +\sqrt{1/12}\ket{\overline{\mbox{ok}}}_{\overline{L}}\ket{\mbox{\tiny ``$\overline{w} = \overline{\mbox{ok}}$, so I am certain that $w = $ fail at $n:31$''}}_{\overline{W}}\ket{\mbox{ok}}_{L}\ket{\tiny\mbox{``I am certain that $w =$ fail at $n:31$''}}_{W}  \nonumber \\
&& + \sqrt{1/3}\ket{\overline{\mbox{fail}}}_{\overline{L}}\ket{\mbox{\tiny ``$\overline{w} = \overline{\mbox{fail}}$; no conclusion drawn''}}_{\overline{W}}\ket{\mbox{fail}}_{L}\ket{\tiny\mbox{``no conclusion drawn''}}_{W} \nonumber \\
&& +\sqrt{1/3}\ket{\overline{\mbox{fail}}}_{\overline{L}}\ket{\mbox{\tiny ``$\overline{w} = \overline{\mbox{fail}}$; no conclusion drawn''}}_{\overline{W}}\ket{\mbox{ok}}_{L}\ket{\tiny\mbox{``no conclusion drawn''}}_{W}  
\end{eqnarray}

Finally, after $W$'s measurement, the global state is
\begin{eqnarray}
\ket{\Psi}^{n:31} & = &  \sqrt{1/12}\ket{\overline{\mbox{fail}}}_{\overline{L}}\ket{\mbox{\tiny ``$\overline{w} = \overline{\mbox{fail}}$; no conclusion drawn''}}_{\overline{W}}\ket{\mbox{fail}}_{L}\ket{\mbox{\tiny ``no conclusion drawn previously; now I observe $w = $ fail''}}_{W}\nonumber \\
&&-\sqrt{1/12}\ket{\overline{\mbox{fail}}}_{\overline{L}}\ket{\mbox{\tiny ``$\overline{w} = \overline{\mbox{fail}}$; no conclusion drawn''}}_{\overline{W}}\ket{\mbox{ok}}_{L} \ket{\mbox{\tiny ``no conclusion drawn previously; now I observe $w = $ ok''}}_{W}\nonumber \\ 
&&-\sqrt{1/12}\ket{\overline{\mbox{ok}}}_{\overline{L}}\ket{\mbox{\tiny ``$\overline{w} = \overline{\mbox{ok}}$, so I am certain that $w = $ fail at $n:31$''}}_{\overline{W}}\ket{\mbox{fail}}_{L}\ket{\mbox{\tiny ``I am certain that $w =$ fail at $n:31$; now I observe $w = $ fail''}}_{W} \nonumber \\
&&+\sqrt{1/12}\ket{\overline{\mbox{ok}}}_{\overline{L}}\ket{\mbox{\tiny ``$\overline{w} = \overline{\mbox{ok}}$, so I am certain that $w = $ fail at $n:31$''}}_{\overline{W}}\ket{\mbox{ok}}_{L}\ket{\mbox{\tiny ``I am certain that $w =$ fail at $n:31$; now I observe $w = $ ok''}}_{W} \nonumber \\
&&+\sqrt{1/3}\ket{\overline{\mbox{fail}}}_{\overline{L}}\ket{\mbox{\tiny ``$\overline{w} = \overline{\mbox{fail}}$; no conclusion drawn''}}_{\overline{W}}\ket{\mbox{fail}}_{L}\ket{\mbox{\tiny ``no conclusion drawn previously; now I observe $w = $ fail''}}_{W}\nonumber \\
&&+\sqrt{1/3}\ket{\overline{\mbox{fail}}}_{\overline{L}}\ket{\mbox{\tiny ``$\overline{w} = \overline{\mbox{fail}}$; no conclusion drawn''}}_{\overline{W}}\ket{\mbox{ok}}_{L}\ket{\mbox{\tiny ``no conclusion drawn previously; now I observe $w = $ ok''}}_{W} \label{finalstate}
\end{eqnarray}

The rules about `certainty' license time-stamped entries in the memory registers of the agents. According to the  state $\ket{\Psi}^{n:31}$ (see the fourth component), for each round $n$ there is a finite  probability, specifically 1/12, that $W$'s internal memory registers $w$ = fail as well as the observation of the measurement outcome $w$ = ok. With probability 1 this will occur for some round. The argument shows that the rules are inconsistent: applying rules $Q$ and $C$ to the Gedankenexperiment leads to a violation of rule $S$.

One might object that the inconsistency here is not a logical contradiction, because that would involve the demonstration of statements $p$ and not-$p$ from $Q$ and $C$, which is not the same thing as deriving `I am certain that $p$' and `I am certain that not-$p$.'  Rule $S$ excludes such conflicting modal statements about certainty, but whether modal statements are contradictory depends on the system of modal logic. This objection misses the point. If we think of the agents as quantum automata or quantum computers, the rules about `certainty' simply justify timed entries in the memory registers. Rule $S$ prohibits conflicting entries with the same time stamp, and the derivation of a conflict for a particular experimental scenario provides a counterexample to the possibility of applying quantum theory  `to model complex systems that include agents who are themselves using quantum theory.' 

It seems that we need to drop at least one of the three inference rules, $Q, C$, or $S$ to avoid this conclusion. Alternatively---and this is my own view---quantum mechanics, as a noncommutative or non-Boolean theory, unlike a classical (Boolean) theory, is not the sort of theory that can be used to give a `view from nowhere' explanation of physical phenomena. Quantum mechanics cannot be applied to the observational processes of observers who are themselves observing other systems. As Asher Peres put it \cite[p. 173]{Peres1995}:
\begin{quote}
While quantum theory can in principle describe \emph{anything}, a quantum description cannot include \emph{everything}. In every physical situation \emph{something} must remain unanalyzed. 
\end{quote}

\section*{Acknowledgements}
Thanks to Renato Renner for helpful discussions, and for making available his `Notes on the Discussion of
``Quantum theory cannot consistently describe the use of itself.'''

\end{document}